\documentclass[
    twocolumn,			
    superscriptaddress, 
    amsmath, 			
    amssymb, 			
    prl,				
    aps,				
    floatfix			
    ]{revtex4-2}

\usepackage{graphicx}	
\usepackage{braket}		
\usepackage{amsmath}    
\usepackage{dcolumn}	
\usepackage{bm}			
\usepackage{subdepth}	
\usepackage{gensymb}	
\usepackage{color,soul}

\begin{document}

\title{Picosecond Pulsed Squeezing in Thin-Film Lithium Niobate Strip-Loaded Waveguides at Telecommunication Wavelengths}%
 
\author{Daniel Peace}
\affiliation{Centre for Quantum Computation and Communication Technology (Australian Research Council), Centre for Quantum Dynamics, Griffith University, Brisbane, QLD 4111, Australia}
\author{Alexander Zappacosta}
\affiliation{Centre for Quantum Computation and Communication Technology (Australian Research Council), Centre for Quantum Dynamics, Griffith University, Brisbane, QLD 4111, Australia}
\author{Robert Cernansky}
\affiliation{Centre for Quantum Computation and Communication Technology (Australian Research Council), Centre for Quantum Dynamics, Griffith University, Brisbane, QLD 4111, Australia}
\affiliation{Institute for Quantum Optics and IQST, Ulm University, Albert-Einstein-Allee 11, D-89081 Ulm, Germany}
\author{Ben Haylock}
\affiliation{Centre for Quantum Computation and Communication Technology (Australian Research Council), Centre for Quantum Dynamics, Griffith University, Brisbane, QLD 4111, Australia}
\affiliation{Institute for Photonics and Quantum Sciences, SUPA, Heriot-Watt University, Edinburgh EH14 4AS, United Kingdom}
\author{Andreas Boes}
\affiliation{Integrated Photonics and Applications Centre (InPAC), School of Engineering, RMIT University, Melbourne VIC 3001, Australia}
\author{Arnan Mitchell}
\affiliation{Integrated Photonics and Applications Centre (InPAC), School of Engineering, RMIT University, Melbourne VIC 3001, Australia}
\author{Mirko Lobino}
\affiliation{Centre for Quantum Computation and Communication Technology (Australian Research Council), Centre for Quantum Dynamics, Griffith University, Brisbane, QLD 4111, Australia}
\affiliation{Department of Industrial Engineering, University of Trento, via Sommarive 9, 38123 Povo, Trento, Italy}
\email{mirko.lobino@unitn.it}
\date{\today}
\begin{abstract}
Achieving high level of pulsed squeezing, in a platform which offers integration and stability, is a key requirement for continuous-variable quantum information processing. Typically highly squeezed states are achieved with narrow band optical cavities and bulk crystals, limiting scalability. Using single-pass parametric down conversion in an integrated optical device, we demonstrate quadrature squeezing of picosecond pulses in a thin-film lithium niobate strip-loaded waveguide. For on-chip peak powers of less than 0.3~W, we measure up to -0.33$\pm0.07$~dB of squeezing with an inferred on-chip value of -1.7$\pm$0.4~dB. This work highlights the potential of the strip-loaded waveguide platform for broadband squeezing applications and the development of photonic quantum technologies. 
\end{abstract}

\maketitle

\section{Introduction}
Squeezed vacuum is a key resource state for quantum protocols based on continuous-variables (CV) \cite{Braunstein2005}. In these protocols the quantum information is encoded using the electromagnetic field quadrature as degree of freedom which has a continuous spectrum. Squeezed states are essential for CV measurement-based quantum computation (MBQC), in which operations are performed through a series of local single-qubit measurements on a large entangled resource state, known as a cluster state \cite{Raussendorf2001}. When combined with time-domain multiplexing protocols \cite{Menicucci2011}, squeezed vacuum has enabled the deterministic generation of the largest cluster states created, containing thousands to millions of entangled modes \cite{Yokoyama2013,Larsen2019,Asavanant2019}.

A broadband source of squeezed vacuum is important for time multiplexed CV-MBQC schemes since the bandwidth defines the minimum width of the temporal mode, while high level of squeezing is required to reach the fault tolerance threshold for error correction \cite{Menicucci2013,Fukui2018,Takeda2019}. Traditionally, squeezing has been generated in optical parametric oscillators, where -15~dB has been measured from a bulk monolithic cavity design \cite{Vahlbruch2016}. Such implementations are however inherently not scalable and have bandwidths limited to megahertz and gigahertz ranges due to their cavity nature. 

A solution to the limitation of bulk cavities comes from photonic integrated circuits, where the strength of nonlinear interactions is increased with tight mode confinement. Platforms such as silicon nitride (SiN) have been used for generating squeezing of up to -1.5~dB with four-wave mixing (FWM)\cite{Vaidya2020, Zhao2020, Cernansky2020}. However, the squeezing bandwidth is again limited due to the weaker third-order $(\chi^{3})$ process requiring microring cavities to enhance the nonlinear interaction. 

Second order nonlinear materials offer stronger nonlinear interactions and, among them, lithium niobate (LN) is already ubiquitous within quantum optics owing to its wide optical transparency window (400~nm - 5~$\mu$m) and large second-order nonlinear susceptibility ($\chi_{33}^2 = 30~\text{pm/V}$). Single-pass optical parametric amplifiers (OPA) with periodically poled lithium niobate (PPLN) waveguides provide a path towards achieving high squeezing with terahertz level bandwidth. Previous demonstrations have shown squeezing levels below -6~dB across 2.5~THz \cite{Kashiwazaki2020} and more recently 6~THz bandwidths \cite{Kashiwazaki2021}. These demonstrations utilised ZnO-doped PPLN waveguides bonded to a LiTaO3 substrate, fabricated using mechanical dicing. They require over 300~mW continuous-wave (CW) pump power given a device length of 45~mm and are fabricated with a technology that is not compatible with bends, therefore not suitable for the integration of other components like directional couplers \cite{Lenzini2018}.


Thin-film lithium niobate (TFLN) has emerged as a leading material platform for quantum optics given its demonstrated ultra-low propagation losses (3~dB/m) \cite{Zhang2017} and large nonlinear efficiencies (4600~\%/Wcm$^2$) \cite{Rao2019}. Ultra-broadband quadrature squeezing has been generated in Z-cut ridge waveguide in TFLN, where -0.56~dB of squeezing across 7~THz bandwidth was measured \cite{Chen2021}. In X-cut TFLN, -4.2~dB squeezing was measured at the 2~$\mu$m wavelength across 25~THz, using femtosecond pulses \cite{Nehra2022}. 

In spite of this, due to the fabrication difficulties associated with ridge-based waveguides propagation losses are typically higher around 0.5~dB/cm \cite{Chen2021} and phase matching is highly sensitive to variations in waveguide width \cite{Wang2018, Boes2019}. These drawbacks can be reduced by using strip-loaded waveguides which avoid etching the LN layer entirely, instead localising the propagating mode within the thin-film through deposition of high refractive index material such as silicon (Si) or SiN on top of the LN layer \cite{Boes2019}. Strip-loaded waveguides achieve comparable propagation losses of $\sim$0.3~dB/cm while maintaining bending radii of 300~$\mu$m\cite{Ahmed2019} and high nonlinear conversion efficiencies. Second harmonic conversion efficiencies of $\sim$1160~\%/Wcm$^2$ have been reported, demonstrating that the guided mode is still largely confined within the LN layer \cite{Chang2016, Boes2019}.

In this work, we generate quadrature squeezing using picosecond pulses in TFLN strip-loaded waveguide at  1556.55~nm. For an average pump power of 310~$\mu$W, we measure a squeezing and anti-squeezing level of -0.33$\pm$0.07~dB and 0.48$\pm$0.06~dB respectively, with an inferred on-chip squeezing of -1.7$\pm$0.4~dB. The telecommunication wavelength and well defined time-bin windows provided by the short optical pulses are useful for a broad range of CV quantum photonic applications, including quantum computation \cite{Takeda2019, Su2018, VanLoock2000}.

\section{Waveguide Design and Properties}
Strip-loaded waveguides have recently been shown to be more susceptible to lateral leakage when compared to ridge waveguides \cite{Boes2019}. This effect can lead to an increase in propagation losses, with shorter wavelengths affected more. For second harmonic generation (SHG) this tends to increase propagation loss of the higher frequency mode, resulting in lower measured efficiency \cite{Chang2016}, or in the context of parametric down conversion, a reduction in overall gain due to the additional loss of pump power. In this experiment we use waveguides engineered such that they do not suffer from any lateral leakage at the pump wavelength \cite{Boes2019}. Our waveguides were fabricated on a TFLN wafer with an X-cut LN film of 300~nm thickness and a 400~nm SiN layer deposited on top \cite{Boes2019}. Waveguides are 2~$\mu$m wide and defined by etching 380~nm of the SiN layer, followed by the deposition of SiO$_2$ as a top cladding. Electric field periodic poling is performed before the deposition of the SIN layer in a regions defined by a series of segmented electrodes patterned on the LN surface. Waveguide facets are diced and polished to maximise end-fire coupling efficiency.

\begin{figure}	
	\centering
	\includegraphics[width=1.0\linewidth]{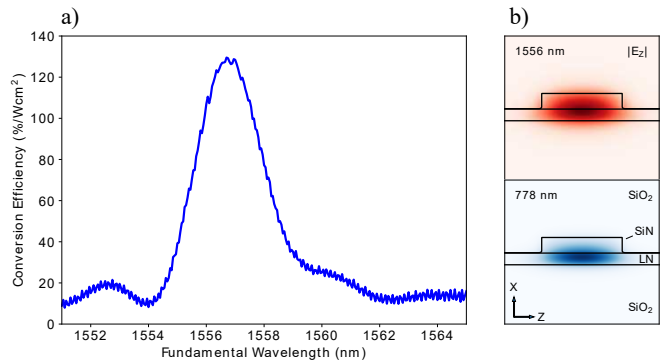}
	\caption{Nonlinear conversion efficiency. (a) SHG conversion efficiency and phase matching bandwidth measured with CW tunable laser. Waveguide demonstrates a maximum normalised conversion efficiency of 127~$\%$/Wcm$^2$. (b) Simulated waveguide mode profiles for the fundamental and SH modes.}
	\label{fig:shg}
\end{figure}

A poling period of 4.93~$\mu$m is used for a phase matching wavelength around 1550~nm with a theoretical conversion efficiency of 1070~\%/Wcm$^2$ \cite{Boes2019}. Waveguide temperature is held to a constant $29^{\circ}\text{C}$, maximising parametric gain. Measured normalised conversion efficiency is 127~\%/Wcm$^2$, equivalent to 28~$\%/$W with a device length of 4.7~mm. Differences between measured and theoretical efficiencies are attributed to defects in the periodic poling, demonstrated by the asymmetry of the SHG generation curve shown in Fig~\ref{fig:shg}. The phase matching spectrum is measured using a tunable CW laser (Tunics Plus CL) with an on-chip power of approximately 2~mW.

When pumping with pulses it is important to consider that waveguides are more dispersive than bulk material due to mode confinement, therefore temporal walk-off must be quantified for efficient nonlinear interactions. Temporal walk-off is given by the accumulated group velocity mismatch such that $\tau_{\text{walk-off}} = \Delta k' L$, where $\Delta k' = v_{g, 2\omega}^{-1} - v_{g, \omega}^{-1}$ is the group velocity mismatch between the SH and fundamental modes, and $L$ the device length\cite{Jankowski2021}. Performing modal analysis at the wavelengths involved, we calculate the total temporal walk-off across the waveguide to be $\tau_{\text{walk-off}}=1.47$~ps. The small temporal walk-off is confirmed in the similarity between CW and pulsed SHG conversion efficiencies.
\begin{figure*}[t]
	\centering
	\includegraphics[width=0.9\linewidth]{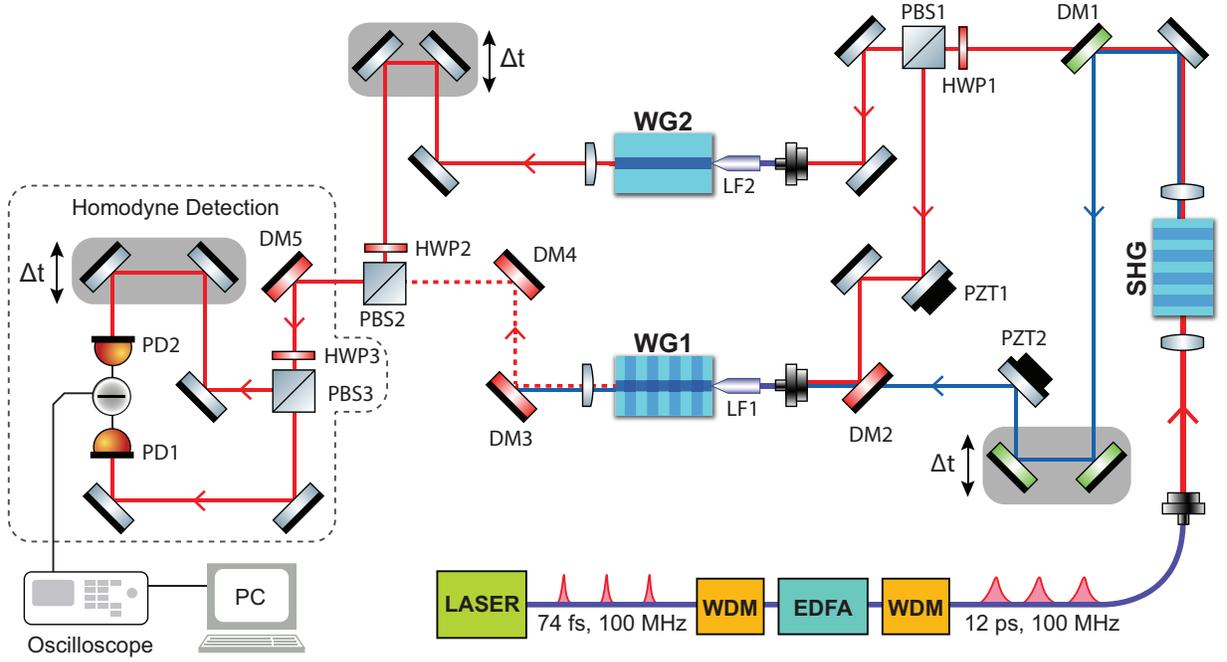}
	\caption{Experimental setup. Red (blue) indicate 1556.6~nm (778.3~nm) beam paths and purple showing optical fibres. Dotted red is the squeezed vacuum beam. Seed path is kept fixed, with time-delay stages in pump and LO paths used for pulse temporal overlapping. LO: local oscillator, WG1: waveguide for the generation of squeezing and gain measurements, WG2: waveguide for the spatial mode filtering of the LO, DM: dichroic mirror, LF: lensed fibre, HWP: half-wave plate, PBS: polarising beam splitter, DWDM: dense wavelength division multiplexer, EDFA: Erbium-doped fibre amplifier, PZT: piezoelectric mirror, PD: photodiode.}
	\label{fig:experiment}
\end{figure*}

\section{Experimental Results}
The experimental setup shown in Fig.~\ref{fig:experiment} was used for the parametric gain and squeezing measurements. The master laser is a pulsed laser (NKT Origami-15 LP) which generates 74~fs $\text{sech}^2$, transform-limited pulses at a 100~MHz repetition rate. Picosecond pulses are generated from the master laser by filtering its bandwidth with a dense wavelength division multiplexing (DWDM) module from which we select a 100~GHz portion of the spectrum from one of its output channels. The pulses are then amplified with an erbium-doped fibre amplifier (EDFA) (Keopsys CEFA-PB-HP). After amplification, we couple the light in a second identical DWDM module to filter out spontaneous emission from the amplifier. Additionally the filtering ensures the pulse spectrum remains within the phase matching bandwidth of the squeezer, assisting the overlap between squeezed and LO pulses in homodyne detection. The resulting pulse has a 12~ps FWHM duration with a central wavelength of 1556.6~nm. 

The pump beam for the generation of squeezing and parametric gain is obtained via SHG from a temperature controlled bulk MgO:PPLN crystal (Covesion MSHG-1550-0.5-40). We generate up to 4~mW average power at 778.3~nm, with a pulse duration of $\sim$10~ps, for a total conversion efficiency of nearly 18$\%$. The pulses at the two wavelengths (1556.6~nm and 778.3~nm) are separated using a dichroic mirror (DM1), with additional filtering added to the pump path to ensure complete removal of the infrared radiation. While most of the power at 1556.6~nm is used for homodyne detection as local oscillator (LO), a small portion is recombined with the pump beam (DM2) and used as seed beam for classical gain measurement. As shown in Fig.~\ref{fig:experiment}, the LO mode is spatially filtered in a second identical waveguide device (WG2) in order to increase the spatial overlap with the squeezed light, resulting in visibility up to 92$\pm$1\% between 1556.6~nm pulses.

Temporal overlap of the three beams (pump, LO and seed) is achieved by keeping the seed beam fixed and adjusting two translation stages located in the LO and pump paths. A third tunable time-delay is located in one arm of the homodyne detector (HD) for optimal temporal overlap of the two photocurrents from the photodiodes. Both waveguides (WG1 and WG2 in Fig.~\ref{fig:experiment}) are coupled to with SMF-28 lensed fibres (OZOptics) designed with a spot size of 2~$\mu$m. An anti-reflection coating is used for the LO fibre to minimise back reflections. Outcoupling from both chips is by identical high numerical aperture lenses (Thorlabs C660TME-C, NA=0.6), achieving an  out coupling efficiency of -3.5~dB. After the squeezer pump pulses are filtered out through several dichroic mirrors (DM3-DM5) between the chip and the homodyne detection, accounting for at least 60~dB of suppression. 

To characterize the performances of the nonlinear waveguide we first measure the parametric gain defined by $G_{\pm} \text{[dB]} = 10 \log_{10}\left( P_{\omega}(L) / P_{\omega}(0) \right)$. The measurement is performed combining the pump beam with a weak seed beam at 1556.6~nm and  applying a ramp function to a piezoelectric transducer (PZT2) to scan the relative between pump and seed by a full $2\pi$. The parametric amplification and deamplification of the seed beam is measured using fixed gain photodiode (Thorlabs PDA015C) placed in one arm of the HD (Fig.~\ref{fig:experiment}). In order to avoid photodiode saturation and pump depletion, the seed power is kept below 2~$\mu$W average power.

\begin{figure}
	\centering
	\includegraphics[width=0.7\linewidth]{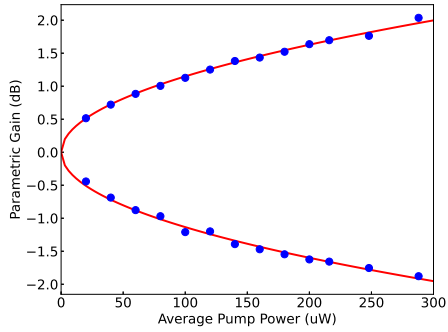}
	\caption{Classical parametric gain versus on-chip pump power for an average seed power of 2~$\mu$W. Standard error for each measurement lies within markers.}
	\label{fig:gaincurve}
\end{figure}

For an on-chip average power of 290~$\mu$W, we achieve up to -1.88$\pm$0.02~dB deamplifcation and 2.04$\pm$0.02~dB amplification. Gain is measured over a range of pump powers as shown in Fig.~\ref{fig:gaincurve}. Fitting the data points with the equation $G_{\pm}=\eta_{mm}\times\text{exp}(\pm 2\sqrt{\alpha P})+1-\eta_{mm}$ \cite{Eto2008}, we estimate the degree of mode-matching between pump and seed pulses to be 95$\pm$3\%.

Squeezing measurements are performed using time-domain balanced homodyne detection (HD). Detector design is based on ref.~\cite{Kumar2012}, featuring InGaAs photodiodes with a quantum efficiency of 98\% quantum at 1550~nm (Laser Components Germany GmbH), 3-dB bandwidth of 80~MHz and a common mode rejection ratio of 55~dB. On the homodyne signal output, two 100~MHz notch filters are used to remove frequency components at the laser repetition rate. Low frequency contributions are removed by a high-pass filter with cutoff frequency of 35~MHz. The homodyne current is digitised via an oscilloscope (Tektronix MSO5204B) with a 1~Gsps sampling rate, recording 5 million points per trace, accounting for 500,000 pulses. Individual data points are first grouped into their respective pulse window and integrated to obtain the quadrature measurement \cite{Kumar2012}. To reconstruct the phase modulation of the PZT, we group 5000 consecutive pulses into a single phase bin.
For each squeezing measurement the calculated variance, $\langle \Delta^2 \hat{X} \rangle$ is averaged across 18 traces. For all squeezing measurements the local oscillator (LO) power is set to an average power of 1~mW, resulting in a shot noise clearance of 8~dB from the electronic noise. 

\begin{figure}
	\centering
	\includegraphics[width=1.0\linewidth]{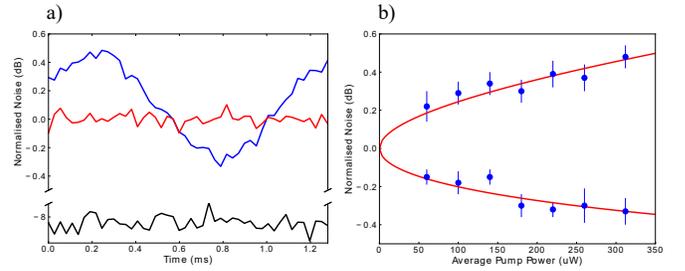}
	\caption{Pulsed squeezing measurement. (a) Normalised electronic noise (black), shot noise (red), and squeezed vacuum noise (blue) as a function of the phase for on chip average power of 310~$\mu$W. Electronic noise  (b) Measured squeezing and anti-squeezing as a function of coupled pump peak power. Error bars represent standard error.}
	\label{fig:squeezingtrace}
\end{figure}

Figure~\ref{fig:squeezingtrace}(a), displays the reconstructed time-domain trace for maximum squeezing and anti-squeezing level achieved and HD clearance. For a maximum on chip average pump power of 310~$\mu$W, we measure a maximum squeezing and anti-squeezing level of $-0.33\pm0.07$~dB and $0.48\pm0.06$~dB. Error values represent the standard error. Squeezing level versus pump power is shown in Fig.~\ref{fig:squeezingtrace}(b), with data points fit using the equation $S_{\pm} =  \eta \times \text{exp} (\pm 2\sqrt{\alpha P}) + 1 - \eta$ \cite{Kashiwazaki2020}, where $\eta$ is the total detection efficiency, $\alpha$ the SHG conversion efficiency and $P$ the peak pump power in the waveguide, we estimate a total detection efficiency $\eta$ of 22$\pm$4\%.

Total detection efficiency $\eta$ is the product of the  transmission through waveguides $\eta_{waveguide}$, transmission of all other optical components $\eta_{ol}$, and HD efficiency $\eta_{HD}$ (which includes quantum efficiency of the photodiodes $\eta_{qe}$, electronic noise $\eta_{e}$, and temporal and spatial overlap of the squeezed vacuum with LO $\eta_{LO}$) \cite{Eto2008}.

 HD clearance contributes an effective loss of $\eta_{e} = 1 -  \langle \Delta^2 \hat{X}_{\text{e}} \rangle / \langle \Delta^2 \hat{X}_{\text{LO}}\rangle$\cite{Kumar2012}, where $\langle \Delta^2 \hat{X}_{\text{e}} \rangle$ and $\langle \Delta^2 \hat{X}_{\text{LO}}\rangle$ represent the variance of electronic and shot noises respectively. Total losses contributed by the HD is therefore, $\eta_{HD}=\eta_{LO}\times\eta_{qe} \times\eta_{e} = 0.85 \times 0.98 \times 0.84 \approx 0.7$, equivalent to -1.55~dB loss. Waveguide losses are estimated to be -0.29~dB based on 0.6~dB/cm propagation losses \cite{Boes2019}. Optical components account for -4.57~dB in detection efficiency, given out coupling and path losses, therefore expected total detection efficiency based on known loss is 23\%, matching closely to the estimated total detection efficiency of $22\pm4$\%. Based on curve fitting and accounting for optical losses outside the waveguide we infer and on-chip squeezing of -1.7$\pm$0.4~dB.
 
 Given reported SHG efficiencies of \cite{Boes2019}, generation of over -10~dB on-chip squeezing is possible in a 5~mm long device assuming a peak power of 1.0~W. Furthermore using domain engineering, quasi-static interactions can further improve squeezing bandwidth to many terahertz wide \cite{Jankowski2021, Ledezma2021, Nehra2022}.

\section{Conclusion}
In summary, we successfully demonstrate quadrature squeezing of picosecond pulses using TFLN strip-loaded waveguides. We measure up to -$0.33\pm0.07$~dB squeezing with an inferred on-chip squeezing level of -1.7$\pm$0.4~dB. This work highlights the potential of strip-loaded LN waveguides for nonlinear optics and as a platform for the integration of other components for optical quantum computing. The demonstrated performance paves a clear path to meeting squeezing requirements for fault-tolerant CV-MBQC. Beyond universal computation the integrated nature is ideal for other CV-QIP applications such as Gaussian boson sampling \cite{Su2018}, or quantum teleportation networks \cite{VanLoock2000}.

\section*{Acknowledgements}
\begin{acknowledgments}
This work was supported by the Australian Research Council (ARC) Centre of Excellence for Quantum Computation and Communication Technology (CE170100012), and the Griffith University Research Infrastructure Program. ML was supported by the Australian Research Council (ARC) Future Fellowship (FT180100055). BH was supported by the Griffith University Postdoctoral Fellowship. This work was performed in part at the Queensland node of the Australian National Fabrication Facility, a company established under the National Collaborative Research Infrastructure Strategy to provide nano- and microfabrication facilities for Australia’s researchers. We thank Stefan Morley for his support with the electronics and the realisation of the homodyne detector.
\end{acknowledgments}

\section*{Data Availability}
The data that support the findings of this study are available from the corresponding author upon reasonable request.

\bibliography{FinalBibliography.bib}

\end{document}